\documentclass[a4paper]{article}
\usepackage{amsmath,amsthm,amssymb, mathrsfs}

\usepackage{fullpage}
\usepackage{enumerate}
 
\usepackage{listings}
\usepackage{tikz-cd}
\usepackage{tikz}
\usepackage{tocbibind}
\usepackage{comment}
\usepackage{mathtools}
\usepackage[english]{babel}
\usepackage[autostyle]{csquotes}
\usepackage[
backend=biber,
style=ieee]{biblatex}

\usepackage{hyperref}

\newcommand{\Co}{\mathbb{C}}
\newcommand{\Z}{\mathbb{Z}}
\newcommand{\N}{\mathbb{N}}

\def \r {\rightarrow}

\def \lan {\langle}
\def \ran {\rangle}

\def \meet {\wedge}
\def \join {\vee}
\def \De {\Delta}

\newtheorem{theorem}{Theorem}[section]
\newtheorem{defin}[theorem]{Definition}
\newtheorem{lemma}[theorem]{Lemma}
\newtheorem{coro}[theorem]{Corollary}
\newtheorem{exmp}[theorem]{Example}
\newtheorem{remark}[theorem]{Remark}
\newtheorem{prop}[theorem]{Proposition}

\newtheorem{no_count_theorem}{Main Result}

\title{Critical Multi-Cubic Lattices: A Novel Implication Algebra for Infinite Systems of Qudit Gates}
\author{Morrison Turnansky\\ \texttt{mat5ej@virginia.edu}\\ \texttt{University of Virginia} \\ 
}
\begin{document}
\maketitle
\begin{abstract}
We introduce a new structure, the critical multi-cubic lattice. Notably the critical multi-cubic lattice is the first true generalization of the cubic lattice to higher dimensional spaces. We then introduce the notion of a homomorphism in the category of critical multi-cubic lattices, compute its automorphism group, and construct a Hilbert space over which we represent the group. With this unitary representation, we re-derive the generalized Pauli matrices common in quantum computation while also defining an algebraic framework for an infinite system of qudits. We also briefly explore the critical multi-cubic lattice as a novel implication algebra serving as a logical framework for qudit gates.  \newline 
\end{abstract}

\section*{Acknowledgements}
\thispagestyle{empty}
\markboth{Acknowledgements}{Acknowledgements}
I would like to thank Professor B. Hayes for the many discussions about this paper and the thesis on which it is based. Also, I am grateful to Dr. J. S. Oliveira for introducing me to the topic of multi-cubic lattices.  

\section*{Statements and Declarations}
No funds, grants, or other support was received.
\section*{Data Availability Statement}
No datasets were generated or analysed.
\newpage
\section{Introduction}
The \hyperlink{multicubic lattice definition}{multi-cubic lattice} \cite{MCL} has long been seen as a successor to the cubic lattice \cite{oliveira} because it maintains many of the qualities of the cubic lattice. However, the multi-cubic lattice is not really a generalization as there is no set of parameters such that a multi-cubic lattice is a cubic lattice. In section 2), we introduce a proper generalization of the cubic lattice, namely the \hyperlink{definition critical multi-cubic lattice}{critical multi-cubic lattice}. Our intuitive inspiration for such a construction is as follows. Just as the cubic lattice can be viewed as the lattice of the faces of a $n$ dimensional cube with an operator, $\De$, representing antipodal symmetry, the critical multi-cubic lattice can be viewed as the lattice of the faces of an $n$ dimensional square grid with operators preserving translational symmetry. As a result of this formulation, we are able to generalize the results of \cite{cubic}. The cubic lattice and its automorphism group were seen to be similar to the spin $\frac{1}{2}$ system of an arbitrary cardinal of qubits, so it is reasonable to consider higher order spin systems as we generalize the results of the cubic lattice.\\
Section 3) begins by defining the morphisms in the category of critical multi-cubic lattices. In the sense that the critical multi-cubic lattice is the generalization of the antipodal symmetry of a cube, an action of $Aut(\Z_{3}) \cong \Z_2$, we have that $Aut(\Z_{2k+1})$ enforces this new symmetry. 

\begin{no_count_theorem}[Theorem \ref{isomorphism of permutations and centralizer}]
Let $M$ be an $|I|$-critical multi-cubic lattice over $\mathbb{Z}_{2k+1}$, and $\wr$ denote an unrestricted wreath product. We have that \hyperlink{permutations of coatoms definition}{$Per_{Aut(\Z_{2k+1})}(C_M)$} $\cong$ \hyperlink{centralizer of automorphism definition}{$C_{S_{2k}}(Aut(\Z_{2k+1}))$} $\wr S_I$. Let $Aut(\Z_{2k+1})$ be generated by $\{\sigma_i\}_{i=1}^k$, then $C_{S_{2k}}(Aut(\Z_{2k+1})) = \cap_{i=1}^k C_{S_{2k}}(\sigma_i)$, and $ C_{S_{2k}}(\sigma_i) \cong \Pi_{j=1}^{l_i} (\Z_{j_i} \wr S_{N_{j_i}})$. 
\end{no_count_theorem}

We highlight that this result reduces to the case of the cubic lattice when $2k+1 = 3$, and the group reduces to an infinite version of the Coxeter group $B_n \cong \Z_2 \wr S_n$. As $C_{S_{2k}}(Aut_{2k+1})$ is cyclic if and only if $2k+1$ is prime, we are able to simplify our results significantly for the prime case.  

\begin{no_count_theorem}[Corollary \ref{prime generalizaiton}]
If $2k+1$ is prime, $\Z_{2k} \wr S_I \cong$ \hyperlink{Aut(M) definition}{$Aut(M)$} $\cong Per_{Aut(\Z_{2k+1})}(C_M)$.
\end{no_count_theorem}

\noindent In \cite{cubic}, we embedded a cubic lattice into a specifically constructed Hilbert space, namely a potentially infinite tensor product of $\Co^2$. Section 4) answers the question: are there other similar lattice embeddings into a Hilbert space? 

\begin{no_count_theorem}[Theorem \ref{embedding multi-cubic}]
Let $H$ be Hilbert space constructed as an infinite tensor product for vector spaces of dimension $2k$, $k \in \N$ over an index set $I$. For the Hilbert lattice $HL$, there exists a critical multi-cubic lattice $M$ such that $M \subseteq HL$, and the atoms of $M$ are projections onto subspaces forming an orthonormal basis of $H$.  
\end{no_count_theorem}

\noindent It is worth noting that to our knowledge, no attempt of any analytic structure has been attempted on the multi-cubic lattice, much less something as strong as an embedding into a Hilbert space. As is standard in quantum logic, we now have projections to serve as propositions. The remainder of Section 4) investigates this. 

\begin{no_count_theorem}[Remark \ref{logic summary}]
A critical multi-cubic lattice $M(\meet, \join, \De, \r)$ is an implication algebra, with a weak form of complementation and propositions faithfully embedded as projections onto a Hilbert space.
\end{no_count_theorem}

In section 5), we consider unitary actions on the critical multi-cubic lattice. With proper representation, the automorphisms of the critical multi-cubic lattice can be seen as set of well known quantum gates acting on an infinite set of qudits, which we demonstrate with this example. 
\begin{no_count_theorem}[Example \ref{representation of generalized pauli matrices}]
As noted if $M$ is an $|1|$ critical multi-cubic lattice over $\Z_{3}$, then $U_{2k}$ is the Hadamard matrix up to a normalization constant. If $M$ is an $|1|$ critical multi-cubic lattice over $\Z_{5}$, then
\[U_{2k} = \frac{1}{2}
\begin{bmatrix}
 1 & 1 & 1 & 1\\
 1 & i & -1 & -i \\
 1 & -1 & 1 & -1 \\
 1 & -i & -1 & i \\
\end{bmatrix}, 
X_{2k} = \begin{bmatrix}
    0 & 1 & 0 & 0 \\
    0 & 0 & 1 & 0 \\
    0 & 0 & 0 & 1 \\
    1 & 0 & 0 & 0\\
\end{bmatrix}\text{, and } 
D_{2k} = \begin{bmatrix}
    1 & 0 & 0 & 0 \\
    0 & i & 0 & 0 \\
    0 & 0 & -1 & 0 \\
    0 & 0 & 0 & -i \\
\end{bmatrix}
.\]
In general up to a normalization constant, $\left[U_{2k}\right]_{ij} = \omega_j^{i-1}$ where $\omega_j$ is $j$th element of the $2k$ roots of unity with counterclockwise ordering starting at 1.  We recognize $U_{2k}$ as the quantum Fourier transform, $X_{2k}$ as the shift matrix, and $D_{2k}$ as the clock matrix.
\end{no_count_theorem}

Originally introduced by \cite{nonions}, it is well known that these matrices form the framework of quantum mechanical dynamics in finite dimensions. Furthermore, $X_{2k}$ and $D_{2k}$ are known as Generalized Pauli matrices, which are a representation of the respective Heisenberg-Weyl group \cite{A_Vourdas_2004}.

As further evidence of the utility of this generalization, we are able to generalize that the Pauli matrices form an orthonormal basis of $M_{2}(\Co)$ to infinite systems of qudits.

\begin{no_count_theorem}[Theorem \ref{B(H) multi-cubic isomorphism}]
Let $M$ be an $|I|$ critical multi-cubic lattice over $\Z_{2k+1}$ where $2k+1$ is prime, $C$ be the coatoms of $M$, and $H$ be constructed as in Theorem \ref{embedding multi-cubic}. \\ Then $B(H) = W^\ast(\{$\hyperlink{quantum fourier transform definition}{$U_H$}\hyperlink{projection subscript definition}{$p_c$}$U^\ast_H\}_{c \in C}, \{p_c\}_{c \in C}) = W^\ast(\{$\hyperlink{representation definition}{$\rho_i$}$(C_{S_{2k}}(Aut(\Z_{2k+1})))\}_{i \in I}, \{p_c\}_{c \in C})$ if and only if $2k+ 1$ is prime.
\end{no_count_theorem}

\section{Defining the Critical Multi-Cubic Lattice and Foundational Results}
For the sake of completeness, we first introduce the multi-cubic lattice as it exists in the literature. We then proceed to define the critical multi-cubic lattice as a quotient of the multi-cubic lattice in Definition \ref{definition critical multi-cubic lattice}. Lastly, we conclude with some general properties such as atomisticity, Proposition \ref{atomistic multi-cubic lattice}, and coatomisticty, Proposition \ref{coatomistic multi-cubic lattice}, that will be used in later sections and validate the utility of our newly introduced structure. 
\begin{defin}
Let $\Omega = \{1,2,\dots, n\}$. For each $i \in \Omega$, let $\overline{M}_i = \{-n_i, \dots, 0, \dots, n_i\}$ be a finite $\Z$-module of odd size. We define $\mathscr{V} = \Pi_{i=1}^n \overline{M}_i$. Elements of $\mathscr{V}$ will be denoted by $\vec{v}$ with the $i^{th}$ component $(\vec{v})_i$. Note also that $\overline{M}_i$ being of odd size is equivalent to the property $2m = 0$ implies $m = 0$. \cite{MCL}
\end{defin}

We have now described the set $\mathscr{V}$ over which our lattice will be defined. We begin with defining the poset on $\mathscr{V}$.

\begin{defin}
For $A \subseteq \Omega$ define $X_A = \langle e_i | i \in A \rangle$, the submodule generated by the standard basis indexed by the set $A$. We now have a poset $P = (\{\vec{a} + X_A | \vec{a} \in \mathscr{V}, A \subseteq \Omega\}, \subseteq)$, henceforth $P$ will be known as a multi-cube. \cite{MCL}
\end{defin}

To be clear, when we say the ordering is determined by $\subseteq$, we mean as the poset of the submodule, $\vec{a} + X_A \le \vec{b} + X_B$ if and only if $A \subseteq B$ and $\vec{b} - \vec{a} \in X_B$. 

We now make $P$ into a join lattice and meet semi lattice. We define the partial meet as follows:

\begin{defin}\hypertarget{multicubic lattice definition}
Let $P$ be a multi-cube and define $(\vec{a} + X_A) \meet (\vec{b} + X_B) = \vec{c} + X_{A \cap B}$ if there exists $c \in (\vec{a} + X_A )\cap (\vec{b} + X_B)$ and undefined otherwise. Now let us define $\delta_{\vec{b} \ne \vec{a}} = \{i \in \Omega | b_i \ne a_i\}$, this is analogous to a support function as $\delta_{\vec{a} \ne \vec{0}} = \{i \in \Omega | a_i \ne 0\}$, precisely the non-zero coordinates of $\vec{a}$. We can now turn $P$ into a join lattice by defining join as $(\vec{a} + X_A) \join (\vec{b} + X_B)  = \vec{a} + X_{A \cup B \cup \delta_{\vec{a} \ne \vec{b}}}$. For a given indexing set, $I$, and odd $k \in \N$, we say that $\overline{M}$ is a $I$-multi-cubic lattice over $\mathbb{Z}_k$ when $\overline{M}$ is the lattice  constructed from the multi-cube $P$ for $\mathscr{V} = \Pi_{i\in I} \Z_k$. \cite{MCL}
\end{defin}

Hence $\overline{M}$ when considered as a lattice is only complete under the join operation but not meet. In order to move to a critical mulit-cubic lattice, we first need to define a notion of a critical element. 

\begin{defin}
Let $\overline{M}$ be a multi-cubic lattice. A vector $\vec{x} \in \vec{a} + X_A$ is critical if $x_i \ne 0$ implies that $e_i \notin X_A$. Let the critical elements of $\overline{M}$ be denoted by $\Gamma(\overline{M})$. 
\end{defin}

\begin{prop}
There exists a well defined function $\Gamma: \overline{M} \r \Gamma(\overline{M})$ defined by $\vec{x} \mapsto \inf\{m \in \Gamma(\overline{M}): \vec{x} \le m \}$ \cite{MCL}.
\end{prop}

In order to have the multi-cubic lattice be a true generalization of the cubic lattice, we will need to make some slight modifications. 
We generalize the notion of a signed set from a three valued set $\{ -1, X , 1\}$ to an $2n+1$ valued set $\{-n, -n-1, \dots, -1, X, 1, 2 ,\dots n-1, n\}$. With the mild change of definition, the $0$ of the original multi-cubic lattice is now referred to as $X$. The importance of this will become clear as we move from a Cartesian product of $\Z$ modules to a tensor product of $\Z$ modules. The algebraic structure in terms of scalar multiplication, addition, and multiplication will be of course different, but more importantly consistent with an embedding into a von Neumann algebra of similar structure to \cite{cubic}.

With the above discussion, we remove the ambiguity of non-critical elements. 

\begin{defin}\label{definition critical multi-cubic lattice}\hypertarget{definition critical multi-cubic lattice}{}
We define a critical multi-cubic lattice,  $M$,  as a  multi-cubic lattice, $\overline{M}$, modulo the equivalence relation where for any $m,n \in \overline{M}$, we have $m \sim n$ if and only if $\Gamma(m) - \Gamma(n) \in X_{\sigma(m)}$ and $X_{\sigma(m)} = X_{\sigma(n)}$. From this point forward, we also allow a critical multi-cubic lattice to be a potentially infinite direct product of $\Z$ algebras. If $M$ is a critical multi-cubic lattice indexed by a set $I$ over a ring $\Z_{2k+1}$, we say that $M$ is an $|I|$ critical multi-cubic lattice over $\Z_{2k+1}$.
\end{defin}

For a critical multi-cubic lattice, $M$, if we let $n = \Gamma(m) + X_{\sigma(m)}$, we see that with this new notion of equality, all elements are critical. In addition, we have the property that for $m,n \in M$, $m \ne n$, then $\Gamma(m) - \Gamma(n) \ne 0$.

One can view a critical multi-cubic lattice as a mixture between a direct product of finite $\Z$ algebras, and the lattice of submodules defined by the closure of $\{\{\Z e_i\}_{i \in I}, 0\}$ under meets and joins. By reducing to the equivalence class, we lose addition of the product ring, preserve to some degree both ring multiplication and $\Z$ scalar multiplication, and gain a structure of non-zero valued outcomes with an additional indeterminate value seen in the cubic lattice.

Before going forward we show that our preserved operations are well defined. The essential concept is that two elements in a multi-cubic lattice map to the same equivalence class in a critical multi-cubic lattice if their value of the indices which are neither $0$ nor $X$ are equal. 

\begin{prop}\label{well defined operations multi-cubic lattice}
Let $\overline{M}$ be a multi-cubic lattice over $\Z_{2k+1}$. Then the operations of scalar multiplication by units in the base ring and ring multiplication by units are independent of equivalence class representative in the corresponding critical multi-cubic lattice $M$.
\end{prop}

\begin{proof}
Let $P: \overline{M} \r M$ be the natural projection map, and $m,n \in \overline{M}$ such that $P(m) = P(n)$.  Since $P(m) = P(n)$, $m_i = n_i$ for all $m_i, n_i \ne 0$. For all $m_i =0$, $P_i(m_i) = X$, and similarly for all $n_j = 0$, $P_j(n_j) = X$. As we assumed that all non-zero entries of $m$, $n$ are identical, we now have that $P(m)$ and $P(n)$ have identical nonzero entries and all other values as $X$.

For nonzero divisor scalar multiplication, $c \in Aut(\Z_{2k+1})$. We have $c \cdot m = c \cdot \Pi_{i \in I - \sigma(m)} m_i + X_{\sigma(m)} = \Pi_{i \in I - \sigma(m)} cm_i + X_{\sigma(m)}$, $m_i \in \Z_{2k+1}$, and $c \cdot n = c \cdot \Pi_{i \in I - \sigma(n)} n_i + X_\sigma(n) = \Pi_{i \in I - \sigma(n)} cn_i + X_\sigma(n)$, $n_i \in \Z_{2k+1}$. Then $cm_i = cn_i$ for all $m_i, n_i \ne 0$, so $c P(m) = P(cm) = P(cn) = c P(n)$ as nonzero divisor scalar multiplication does not change which indices had 0 values. 

Similarly we show multiplication of units is well defined. Let $\mu \in \overline{M}$, and further all nonzero induces of $m$ and $\mu$ are units in $\Z_{2k+1}$. As $m_i = n_i$ for all nonzero entries, we have $m_i\mu_i + X_{\sigma(m) \cup \sigma(\mu)} = n_i\mu_i + X_{\sigma(n) \cup \sigma(\mu)}$. Then $P(m\mu) = P(n\mu)$ as they agree on all nonzero entries. 
\end{proof}

The atoms and coatoms will play a significant role in our results, so we offer some intuition. 
\begin{exmp}\label{atoms of critical multic-cubic lattice example}
Let $M$ be a $2$ critical multi-cubic lattice over $\mathbb{Z}_{5}$. The atoms of $M$ are of the form $\{a, b\}$ where $a,b \in \{-2,-1,1,2\}$. This generalizes the two dimensional cubic lattice case, where atoms are of the form $(A^-,A^+)$, $|A^- \cup A^+| = |S| = 2$. We assign each index in $S$ to a either the set $A^-$ or the set $A^+$. We can equivalently assign each index in $S$ to either value $-1$ or the value $1$, and see that this is the reduction to the $2$ critical multi-cubic lattice over $\Z_3$. In general, an $|I|$ critical multi-cubic lattice has $(2k)^{|I|}$ atoms of the form $\Pi_{i \in I} a_i$, $a_i \in \Z_{2k+1} - \{0\}$.
\end{exmp}

\begin{exmp}\label{coatoms of critical multic-cubic lattice example}
Let $M$ be a $2$ critical multi-cubic lattice over $\mathbb{Z}_{5}$. The coatoms of $M$ are of the form $\{a, b\}$ where exactly one $a,b \in \{-2,-1,1,2\}$ and all others are assigned to $X$. This generalizes the two dimensional cubic lattice case, where atoms are of the form $(A^-,A^+)$, $|A^- \cup A^+| = |1|$. We assign exactly one index in $S$ to a either the set $A^-$ or the set $A^+$. We can equivalently assign exactly one index in $S$ to either value $-1$ or the value $1$, and see that this is the reduction to the $2$ critical multi-cubic lattice over $\Z_3$. In general, an $|I|$ critical multi-cubic lattice has $2k|I|$ coatoms of the form $\Pi_{i\in I} V$, where $V = X$ for all but exactly one $i \in I$ and $V \in \Z_{2k+1}- \{0\}$ otherwise. One may find it helpful to think of X as the submodule $\Z_{2k+1}$ on each index. 
\end{exmp}

The fact that the atoms and coatoms in the $2$ critical multi-cubic lattice of $\Z_3$ are the same as the $2$ dimensional cubic lattice is not a coincidence. We show that the case of the cubic lattice is just a specific case of the critical multi-cubic lattice with an automorphism. 

\begin{theorem}\label{generaliztion of cubic lattice}
Let $M$ be a $|S|$-critical multi-cubic lattice over $\mathbb{Z}_3$. Then M is a cubic lattice. 
\end{theorem}

\begin{proof}
We need only show that $M \cong L(S)$ as lattices. For each $m \in M$, we have that $m = \Pi_{i \in I - \sigma(m)} a_i + X_{\sigma(m)}$, $a_i \in \Z_3^*  = \{-1, 1\}$, is mapped to an element $a = (A^+, A^-) \in L(S)$ defined by $i \in A^+$ if $a_i = 1$ and $i \in A^-$ if $a_i = -1$, so that $A^+ \cap A^- = \emptyset$. We denote this mapping by $f:M \r L(S)$, and one can see that f is a bijection.

It remains to show that we have an order homomorphism. If $m \le n$, $f(m) = (A^+ , A^-)$, and $f(n) = (B^+ ,B^-)$, then $m - n  \in X_{\sigma(n)}$,  which implies that $B^+ \subseteq A^+$ and $B^- \subseteq A^-$. Therefore, $f(n) = (B^+ , B^-)  \subseteq (A^+ , A^-) = f(m)$. As order homomorphism induce lattice homomorphism we have that $f$ is also a lattice homomorphism.
\end{proof}

We have shown many similarities between cubic lattices and multi-cubic lattices, we now highlight that they are in fact a weaker object as general critical multi-cubic lattices do not meet the axiomatic definition of cubic lattices. 

\begin{prop}
Let $M$ be a critical multi-cubic lattices over $\Z_{2k+1}$, $k \ge 2$. Then $M$ does not meet axiom 2 of Definition \cite[Definition 1.1.2]{cubic}. 
\end{prop}

\begin{proof}
Let $\phi$ be any order automorphism of $M$. Then for any coatoms $a,b \in M$ such that $a \meet b = 0$ implies $a, b \in \{a_je_i: a_j \in \Z_{2k+1}\}$ for some fixed $i \in I$. Since $\phi$ is an order automorphism, $\phi(b)$ is a coatom as well, and $a \join \phi(b) <1$ implies $\phi(b) = a$ because the join of any two distinct coatoms is equal to 1. 
Then for a fixed a, $\phi(b) = a$ for all $b \in \Z_{2k+1} - \{a, 0\}$, so $\phi$ is not injective. 
\end{proof}

In summation, the failure of a general multi-cubic lattice comes down to the fact that $|\Z_{2k+1} - \{a, 0\}| > 1$ for $k \ge 2$. However, other axiomatic properties of the cubic lattice hold.

\begin{prop}\label{atomistic multi-cubic lattice}
Let $M$ be a critical multi-cubic lattice. Then M is a atomistic.  
\end{prop}

\begin{proof}
Let $M$ be an $|I|$ critical multi-cubic lattice over $\Z_{2k+1}$, and $\Pi_{i \in I - \sigma(m)} a_i + X_{\sigma(m)} = m \in M$, $a_i \in \Z_{2k+1} - \{0\}$. Then $m = \join_{j \in \sigma(m)} m_{j}$, where $m_{j_i} = m_i$ for all $i \in I -\sigma(m)$, $m_{j_j} = -1$, and $m_{j_i} = 1$ for all $i \in \sigma(m) - \{j\}$. We have that each $m_j$ is an atom of $M$, so the result follows. 
\end{proof}

\begin{prop}\label{coatomistic multi-cubic lattice}
Let $M$ be a critical multi-cubic lattice. Then M is a coatomistic. 
\end{prop}

\begin{proof}
As $M$ is a critical multi-cubic lattice, for all $m \in M$, $m_i \in \Z_{2k+1} - \{0\}$ or $m_i = \Z_{2k+1}$. Therefore, for any $m \in M$, we have that $m = \Pi_{i \in I- \sigma(m)} a_i + X_{\sigma(m)}$ where $a_i \in \Z_{2k+1} - \{0\}$, so $m  = \meet_{i \in I} c_i(a)$ where $c_i(a)$ denotes the coatom in the $i$th index equal to $a$. 
\end{proof}

\section{Automorphisms of the Critical Multi-Cubic Lattice}
As is often the case with the introduction of a new structure, one often asks what are the important properties that should be preserved by homomorphisms? This leads to another natural question: given a characterization of homomorphisms, can we classify the respective automorphism group up to isomorphism? These two basic questions will be the focus of this section, and they will allow us to study their unitary representations in later sections.

Informally, a critical multi-cubic homomorphism should preserve any of its well defined operations of which there are four, namely $\meet$, $\join$, scalar multiplication by units of the module, and multiplication of units in the ring. Since order homomorphisms are  lattice homomorphisms, and our lattice is coatomistic, it is necessary that a critical multi-cubic lattice homomorphism maps coatoms to coatoms. In addition, any homomorphism of a critical multi-cubic lattice should also preserve scalar and ring multiplication. We formalize the above in the following definition. 

\begin{defin}
Let $k\in \Z$, $I_M, I_N$ be indexing sets, and $M$ be an $|I_M|$ critical multi-cubic lattice over $\Z_{2k+1}$  with coatoms $C_M$, and $N$ be an $|I_N|$ critical multi-cubic lattice over $\Z_{2k + 1}$ with coatoms $C_N$. A map $\phi: M \r N$ is a critical multi-cubic homomorphism if $\phi$ a scalar multiplication homomorphism over the units of the base ring such that for all $c \in C_M$ either $\phi(c) \subseteq C_N$ or $\phi(c) = 0$ . In particular if $N = M$ and $\phi$ is a $\Z$ module automorphism, then $\phi \in Per(C_M)$.   
\end{defin}

Sufficiency of the conditions follows as we are guaranteed to keep the $\Z$ linear structure, and as a function of coatoms between coatomistic lattices, we have an order homomorphism, which is necessarily a lattice homomorphism as well. In fact, ring multiplication is preserved by any permutation of indices, so we do not include in the definition. We will show that this notion is a generalization of a cubic homomorphism. 

\begin{defin}\hypertarget{Aut(M) definition}{}
Let $k \in \N$ and $M$ be a critical multi-cubic lattice over $\Z_{2k+1}$. We denote $Aut(M)$ as the automorphism group of $M$.
\end{defin}

We show by example that the homomorphism conditions are not equivalent.

\begin{exmp}
Let $M$ be a 1 critical multi-cubic lattice  over $\Z_3$ i.e. a cubic lattice.  Now we demonstrate that there are $\Z$ module automorphisms that are not order automorphisms. 
\[A = \begin{bmatrix}
1 & 1 \\
0 & 1\\
\end{bmatrix}.\] $A[0,1]^t  = [1, 1]^t \notin C_M$, so $A$ is a $\Z$ linear transformation that is not even a permutation of coatoms and therefore not an order homomorphism.

We will show that there are order automorphisms that are not $\Z$ module automorphisms. Let $N$ be a 2 critical multi-cubic lattice over $\Z_3$. The atoms of $N$ are the standard basis. Let $P_{ij}$ denote the projection onto the sum of standard basis vectors $e_i$, $e_j$, so that $C_N = \{P_{12}, P_{13}, P_{13}, P_{34}\}$, and with this ordering let $\sigma = (1234) \in S_4 \cong Per(C_N)$. We claim that $\sigma$ is not representable as an invertible linear transformation. Recall that permutations act on the lattice of projections by inner automorphism. Then   $\sigma(P_{12})\sigma^{-1} = P_{13}$, and $\sigma(P_{34})\sigma^{-1} = P_{12}$, but $\sigma(P_{12} + P_{34}) \sigma^{-1} = \sigma I  \sigma^{-1} = I \ne P_{13} + P_{12} = \sigma P_{12} \sigma^{-1} + \sigma P_{34} \sigma^{-1}$.
\end{exmp}

With this view in mind we have a new and equivalent notion of the poset of a critical multi-cubic lattice. 

\begin{defin}
Let $\Pi_i$ be the coordinate-wise projection onto the ith index and $\Pi_J$ be the projection onto the $J \subseteq I$ coordinates. Implicitly, if $m = \Gamma(m) + X_{\sigma(m)}$, we define $\Pi_J$ as $\Pi_{J}^m$, $J^m \subseteq I - \sigma(m)$ where $J^m =  J - \sigma(m)$. 
\end{defin}

\begin{prop}\label{order by projections}
Let $M$ be a critical multi-cubic lattice and  $m$, $n \in M$ such that $m \le n$. Then there exists $\Pi_J$, $J \subseteq I - \sigma(n)$ such that $\Pi_J(m) = n$. 
\end{prop}

\begin{proof}
Recall that $m \le n$, then $m - n \in X_{\sigma(n)}$, so let $J = \sigma(n) - \sigma(m)$.
\end{proof}
Because we fix a generating set of the $\Z$ algebra, we can view critical multi-cubic homomorphisms as an automorphism composed with a projection as defined above. With the importance of the critical multi-cubic automorphisms now highlighted, we proceed to classify them as a generalization of \cite{turnansky}.
\begin{defin}\hypertarget{permutations of coatoms definition}{}
Let $Per_{Aut(\Z_{2k+1})}(C_M)$ be the permutations of coatoms of M that commute with the action of $Aut(\Z_{2k+1})$ defined by $c \mapsto ac$ for $c \in C_M$ and $a \in Aut(\Z_{2k+1})$.
\end{defin}

\begin{lemma}\label{automorphism isomorphic to permutation}
Let $M$ be an $|I|$ critical multi-cubic lattice over $\mathbb{Z}_{2k+1}$. Then $Aut(M) \cong Per_{Aut(\Z_{2k+1})}(C_M)$.
\end{lemma}

\begin{proof}
By definition, we have an injective group homomorphism $i :Aut(M) \r Per(C_M)$ where $i$ is inclusion map. We want to show the range of $i$ is strictly contained in $Per_{Aut(\Z_{2k+1})}(C_M)$. Any $\phi \in Aut(M)$ is a $\Z$ module homomorphism of $\Z_{2k+1}^I$, so we have that for any $a \in Aut(\Z_{2k+1})$, $a$ induces an automorphism on $\mathbb{Z}_{2k+1}$ by multiplication. Now by the property of module homomorphisms over commutative rings for any $m \in M$, $\phi(a) \circ \phi(m) = a \phi(m) = \phi(am) = \phi(ma) = \phi(m) \circ \phi(a)$. Thus, $Aut(M) \le Per_{Aut(\Z_{2k+1})}(C_M)$. 

For the reverse direction, we show that any $\psi \in Per_{Aut(\Z_{2k+1})}(C_M)$ defines a critical multi-cubic automorphism. Firstly, it is a permutation of coatoms of $M$ by definition, and thus bijective. Secondly, for the linearity condition, let $\psi_a$ be multiplication by $a \in Aut(\Z_{2k+1})$, and $m \in M$. Then $a\psi(m) = \psi_a \circ \psi(m) = \psi \circ \psi_a(m)= \psi(am)$.
\end{proof}

Unlike the case of the cubic lattice, the permutations that commute with $Aut(\Z_{2k+1})$ are in general a larger group than just the automorphism group, so we must consider some additional factors. As we will show, it is for this reason that the automorphism groups of the cubic lattice and critical multi-cubic lattice become more similar when $2k+1$ is prime.  

\begin{defin}\hypertarget{centralizer of automorphism definition}{}
As $Aut(\Z_{2k+1})$ fixes the identity, by relabeling, we can consider the action on $\{a:1 \le a \le 2k\}$ by left multiplication. With this action, let $C_{S_{2k}}(Aut(\Z_{2k+1}))$ denote the centralizer of $Aut(\Z_{2k+1})$ in $S_{2k}$.
\end{defin}

\begin{theorem}\label{isomorphism of permutations and centralizer}
Let $M$ be an $|I|$ critical multi-cubic lattice over $\mathbb{Z}_{2k+1}$, and $\wr$ denote an unrestricted wreath product. Then $Per_{Aut(\Z_{2k+1})}(C_M) \cong C_{S_{2k}}(Aut(\Z_{2k+1})) \wr S_I$,  Let $Aut(\Z_{2k+1})$ be generated by $\{\sigma_i\}_{i=1}^k$, then $C_{S_{2k}}(Aut(\Z_{2k+1})) = \cap_{i=1}^k C_{S_{2k}}(\sigma_i)$, and $ C_{S_{2k}}(\sigma_i) \cong \Pi_{j=1}^{l_i} (\Z_{j_i} \wr S_{N_{j_i}})$. 
\end{theorem}

\begin{proof}
Any $\phi \in Per_{Aut(\Z_{2k+1})}(C_M)$ is determined by the image of its coatoms.  Note that for any two distinct coatoms their meet is empty if and only if they are of the form, $ae_i$ and $be_i$ for distinct $a,b \in \Z_{2k+1} - \{0\}$ and some $i \in S$. By Lemma \ref{automorphism isomorphic to permutation}, $\phi$ defines a critical multi-cubic automorphism, so we have that for any $i \in S$ and distinct $a$, $b \in \Z_{2k+1}- \{0\}$, $\emptyset = \phi(ae_i \meet be_i) = \phi(ae_i) \meet \phi(be_i)$. Hence, for all $\alpha \in \Z_{2k+1} -\{0\}$, $\phi(\alpha e_i) =  \beta_\alpha e_j$ for some $\beta_\alpha \in \Z_{2k+1}- \{0\}$ and a fixed $j \in S$. 

We now need only consider $\phi$ for each individual index.  $Aut(\Z_{2k+1})$ acts on the coatoms of fixed index, $C_i = \{a \in \Z_{2k+1}- \{0\}\}$ by left multiplication, so by relabeling, we consider the action on $\{a:1 \le a \le 2k\}$. Let $Per_{Aut(\Z_{2k+1})}(C_i)$ denote the permutations of $C_i$ commuting with the action of $Aut(\Z_{2k+1})$. We observe that $Per_{Aut(\Z_{2k+1})}(C_i)$ is isomorphic to the centralizer of $Aut(\Z_{2k+1})$ in $S_{2k}$. Thus, $\phi \in C_{S_{2k}}(Aut(\Z_{2k+1})) \wr S_{I}$.

For a given $\sigma \in S_{m}$,  with a standard cycle decomposition consisting of $N_j$ cycles of length $j$, we use that $C_{S_m}(\sigma) \cong \Pi_{j=1}^l (\Z_{j} \wr S_{N_j})$. As the centralizer of subgroup is the intersection of the centralizer of its generators, and we conclude our result.
\end{proof}



It is known that the hyperoctehedral group can be represented as the signed permutation group. We have used this terminology to inspire our generalization. 

\begin{defin}
We define the $Aut(\Z_{2k+1})$-value permutation group of degree $I$ as a generalization of the elements of the permutation group in $M_I(\Z_{2k+1})$ where each nonzero element can take any value of $Aut(\Z_{2k+1})$. Note that $M_I(\cdot)$ denotes the $(\cdot)$ valued matrices indexed by $I$. 
\end{defin}

\begin{prop}\label{wreath product representation}
Let $M$ be an $|I|$ critical multi-cubic lattice over $\Z_{2k+1}$ for $k \in \N$. Then there exists a group representation $\rho: Aut(\Z_{2k+1}) \wr S_I \r M_I(\Z_{2k+1})$ as the $Aut(\Z_{2k+1})$-value permutation group of degree $I$. Furthermore, when considering as the base ring $Aut(\Z_{2k+1}) \subseteq Z(\rho( Aut(\Z_{2k+1}) \wr S_I))$.
\end{prop}

\begin{proof}
We have that the standard basis vectors $\{e_i\}_{i \in I}$ form a generating set of $\Z_{2k+1}^I$ as a $\Z$ module. Now $Aut(\Z_{2k+1}) \wr S_I$ can be represented as the $Aut(\Z_{2k+1})$-value permutation group by considering the action $e_i \mapsto a_ie_\sigma(i)$ for a given $(\times_{i \in I} a_i, \sigma) \in Aut(\Z_{2k+1}) \wr S_I$, which we define as $\rho$. As a commutative base ring, $\Z_{2k+1}^\ast \subseteq Z(M_I(\Z_{2k+1}))$, which are in the $Aut(\Z_{2k+1})$-value permutation group by definition and in $Z(Aut(\Z_{2k+1}) \wr S_I)$ because $\rho$ is a multiplicative group representation. 
\end{proof}

\begin{lemma}\label{wreath product subgroup of automorphism group}
Let $M$ be an $|I|$ critical multi-cubic lattice over $\mathbb{Z}_{2k+1}$, and $\rho$ be defined by Proposition \ref{wreath product representation}. Then $\rho(Aut(\Z_{2k+1}) \wr S_I) \le Aut(M)$.
\end{lemma}

\begin{proof}
We claim that $\rho(Aut(\Z_{2k+1}) \wr S_I) \le Aut(M)$. We have a bijection, $f: \{ae_i:a \in \Z_{2k+1} - \{0\}, \, i \in I\} \r C$, where $e_i$ denotes the standard basis vector, and $C$ is the coatoms of the form specified in Example \ref{coatoms of critical multic-cubic lattice example}, and $f$ is defined by $ae_i \mapsto \Pi_{j \in I} V_j$, $V_i = a$, and $V_j = X$ for all $j \ne i$. Then we have that $\rho(Aut(\Z_{2k+1}) \wr S_I)$ defines a permutation of the coatoms by its action on $\{ae_i:a \in \Z_{2k+1} - \{0\}, \, i \in I\}$. We have already shown in Proposition \ref{wreath product representation} that $\rho(Aut(\Z_{2k+1}) \wr S_I)$ commutes with the action of $\Z$ scalar multiplication of units on $\{ae_i:a \in \Z_{2k+1} - \{0\}, \, i \in I\}$, so we conclude our result.  
\end{proof}

\begin{theorem}
Let $M$ be an $|I|$ critical multi-cubic lattice over $\mathbb{Z}_{2k+1}$. Then $Aut(\Z_{2k+1}) \wr S_I \le Aut(M) \cong Per_{Aut(\Z_{2k+1})}(C_M)$.
\end{theorem}

\begin{proof}
As a result of Lemma \ref{wreath product subgroup of automorphism group} and Lemma \ref{automorphism isomorphic to permutation}, we have $Aut(\Z_{2k+1}) \wr S_I$ is isomorphic to a subgroup of $Per_{Aut(\Z_{2k+1})}(C_M)$.
\end{proof}

We have now rederived the group theoretic results of \cite{turnansky} and shown it in much more generality. We highlight the cyclic case below as the operator algebraic structure will be most similar to the results of Section 5, and identical if $k = 1$.

\begin{coro}\label{prime generalizaiton}
If $2k+1$ is prime, $\Z_{2k} \wr S_I \cong Aut(M) \cong Per_{Aut(\Z_{2k+1})}(C_M)$.
\end{coro}

\begin{proof}
By Theorem \ref{isomorphism of permutations and centralizer}, $Per_{Aut(\Z_{2k+1})}(C_M) \cong C_{S_{2k}}(Aut(\Z_{2k+1})) \wr S_I$. If $2k+1$ is an odd prime, then $Aut(\Z_{2k+1}) \cong \Z_{2k}$, and $C_{S_{2k}}(\Z_{2k}) \cong (e \times e) \times (\Z_{2k} \times e) \cong \Z_{2k}$, where $e$ denotes the group consisting of only the identity.
\end{proof}

\section{Logic of the Critical Multi-Cubic Lattice}
Up the this point, we have explored the algebraic structure of the critical multi-cubic lattice. In order to consider a propositional logic, we require a set of propositions and logical connectives. Since, we are starting with a lattice, we begin with the connectives of $\meet$ and $\join$, and our set of propositions. 

In order to embed our propositions as projections onto a Hilbert space as is standard with quantum logic, we proceed to generalize \cite[Theorem 2.1.1]{cubic} to the critical multi-cubic lattice. In the following theorem, we are using the infinite tensor product detailed in \cite{VN2}.

\begin{theorem}\label{embedding multi-cubic}
Let $H$ be Hilbert space constructed as an infinite tensor product for vector spaces of dimension $2k$, $k \in \N$ over an index set $I$. For the Hilbert lattice $HL$, there exists a critical multi-cubic lattice $M$ such that $M \subseteq HL$, and the atoms of $M$ are projections onto subspaces forming an orthonormal basis of $H$.   
\end{theorem}

\begin{proof}
We see that each simple tensor $\otimes_{i \in I} a_i$, $a_i \in \Z_{2k+1} - \{0\}$,  is a C-sequence \cite{VN2}, and we represent it by a respective projection operator. We use the notation $\Pi_{i \in I} a_i$ to define an element $m$ forming the atoms of $M$ an $|I|$ critical multi-cubic lattice over $\Z_{2k+1}$.

Let $\join_M$ denote join in $M$ and $i: M \r HL$. We need only show that for all $a, b \in M$, $i(a \join_M b) \in HL$. Let $a = \Pi_{i \in I} a_i$ and $b = \Pi_{i\in I} b_i$, then $a \join_M b = \Pi_{i \in J} a_i + X_{I - J}$, where $J = \{i \in I: a_i = b_i\}$, and $i(a \join_M b)$ be the projection $V = \otimes_{i \in I} V_i$,  $V_i = a_i$ for $i \in J$ and $V_i = \Co^{2k}$, $i \in J$. By atomisticity of Proposition \ref{atomistic multi-cubic lattice}, $i$ is an order homomorphism and therefore a lattice homomorphism. 

The set of simple tensors are all rank 1 and therefore atoms in $HL$. The atoms of $M$ form an orthonormal system in $H$. For any distinct atoms $a$, $b \in M$, we have that there exists $i \in I$ such that $a_i \ne b_i$, so $\langle a_i , b_i \rangle_{H_\alpha} = 0$ which implies that $\langle a ,b \rangle_H = 0$. Furthermore, these vectors span $\Pi'\otimes_{\alpha \in I} H_\alpha,$ and therefore are dense in $H$.
\end{proof}




\hypertarget{projection subscript definition}{}To relate to our previous notation, we can see that for all $a \in M$, $a$ can be considered as an element of $p \in HL$ by the tensor coordinatization $p_i  = I_{2k}$ for $i \in \sigma(a)$ and $p_i = p_{a_i}$, the projection onto the $a_i$ subspace of $C^{2k}$, otherwise. 


\begin{remark}
We want to highlight that the lattice homomorphism defined in Theorem \ref{embedding multi-cubic} is defined on the lattice $M$ considered as a subset of $HL$ and not a a lattice homomorphism from $M$ to the lattice $HL$. 
\end{remark}

We have seen that we can embed a critical multi-cubic lattice to a Hilbert lattice in much the same way the cubic lattice to the Hilbert lattice. From an analytic perspective these objects have been shown to share many of the same qualities. However this is where the similarities stop for the most part. By the arguments of the previous section, we have removed the cardinality constraints of \cite{MCL}, and defined and abstracted to critical multi-cubic auotomorphisms.

Now that we have projections acting as propositions, we can explore an additional connective, complement. One of the interesting characteristics of the cubic lattice is that we had an order preserving complementation, $\De$. In fact, the action of $\De$ on the coatoms of cubic lattice embedded in the Hilbert lattice exactly matched the action of negation on the same coatoms. However, we lose this in the general case.

\begin{prop}
The action of $^\perp$ on a critical multi-cubic lattice over $\Z_{2k+1}$, $k >2$, $M$, embedded in Hilbert lattice, $HL$, in the manner of Theorem \ref{embedding multi-cubic} does not define a permutation on $C$, the coatoms of $C$ of $M$. 
\end{prop} 

\begin{proof}
We can show that $^\perp$ does not map coatoms to coatoms. Fix a a coatom $ae_i = c \in C$, then $c^\perp = \join_{b \in \Z_{2k+1}- \{a,0\} } be_i$, which is not even in $M$. 
\end{proof}

Although the action $\perp$ does not act nicely on the critical multi-cubic lattice, we still have an order preserving map that is nearly a complement. We are now in a position to readdress the definition of $\De$ in the context of a critical multi-cubic lattice. We first recall the definition of $\De$ on the multi-cubic lattice.

\begin{defin}
Let $\overline{M}$ be a multi-cubic lattice and $a, b \in M$ and $a \le b$. We define $\sigma(a)$ to be the subset of $\Omega$ $i \notin \sigma(a)$ if and only if $a_i \ne 0$, and $\Delta: \overline{M} \times \overline{M} \rightarrow \overline{M}$ such that $\Delta(b,a) = 2\Gamma(b) - \Gamma(a) + X_{\sigma(a)}$ \cite{MCL}.
\end{defin}

Equivalently, we can define $\De$ purely by its action of the multi-cubic lattice.  

\begin{defin}
Let $\overline{M}$ be a multi-cubic lattice and $a,b \in \overline{M}$, $a \le b$. Then we define $\Delta: \overline{M} \times \overline{M} \rightarrow \overline{M}$ by  
 \[  \Delta(b,a)_i  = \begin{cases} 
          -a_i & a_i \in X_{\sigma(b) -\sigma(a)}\\
          a_i & \text{otherwise}\\
       \end{cases}
    \]
 for all $i \in I.$ 
\end{defin}

As a quick justification, we note that both actions flip the sign of $a_i$ if and only if $\Gamma(a)_i$ is nonzero and $\Gamma(a)_i  \in X_{\sigma(b)}$. Another phrasing is that if a and b are critical with $a \le b$, then the coordinate, i, will change sign exactly when $i \in \{B - A\}$. However, this action defined on product $\Z$ modules will be defined as just another module homomorphism when $b = 1$.

By the definition of $\Delta$, one can directly see that it utilizes the $\Z$ module structure of the MCL. However, we can abstract $\Delta$ by only focusing on its action and define it in the context of a multivalued signed set of the MCL. This will allow us to generalize the ideas proven for the cubic lattice. 

\begin{defin}
Let $M$ be a critical multi-cubic lattice over $\Z_{2k+1}$. For all $n \in M$, we define $\De_M(n, \cdot)$ as multiplication by $2k$ on the respective ideal $(n)$.
\end{defin}

\begin{lemma}\label{multi-cubic delta properties}
Let $M$ be a critical multi-cubic lattice, $n \in M$. Then $\De_M(n, \cdot)$ always exists and is well defined.
\end{lemma}

\begin{proof}
For any odd $k \ge 3$, $\De_M$ is just scalar multiplication by $2k$ which exists by construction. As $gcd(2k, 2k+1) = 1$, $\De$ and is well defined by Proposition \ref{well defined operations multi-cubic lattice}.
\end{proof}

The fact that such a map factors through the projection map should not be surprising as the original definition operated on on the equivalence class representatives.
\begin{theorem}
Let $M$ be a critical multi-cubic lattice such that $M \cong L(S)$ for an indexing set $S$ for the lattice isomorphism $f$ of Theorem \ref{generaliztion of cubic lattice}. The action of $\De_M(n,m)$ on $M$ is equal to the action of $\De(f(n), f(m))$ on $L(S)$.
\end{theorem}

\begin{proof}
As the cubic lattice atomistic, we need only show that the claim holds for the atoms of the respective ideal. By the proof of Lemma \ref{multi-cubic delta properties}, we have a well defined $\De_M(n , m)$ that is equivalent to multiplication by $-1$ on the ideal $(n)$. 

When considering the action of the pushforward on $L(S)$, $f \circ \De_M(n,m)$, we swap all nonzero indices of $m \in \sigma(n)$ that are not equal to  $1$ to $-1$ or vice versa. If we let $(B^-, B^+) = f(n)$, and $(A^-, A^+) = f(m)$, then the pushforward is equal to $(B^- \cup \{A^+ - B^+\}, B^+ \cup \{A^- - B^-\}) = \De(f(m), f(n))$.
\end{proof}

While the interplay of $\De$ and $\perp$ is lost, $\De$ has some of the properties of an order preserving compliment.  

\begin{coro}\label{delta properties critical multi-cubic lattice}
Let $M$ be a critical multi-cubic lattice, $a, b \in M$. Then the following hold.
\begin{enumerate}
\item $\Delta(a,a) = a$
\item if $a \le b$, then $\Delta(b,a) \le b$
\item if $a \le b$, then $\Delta(b,a) = 2b - a$
\item if $a \le b$ then $\Delta(b, \Delta(b,a)) = a$
\item if $a \le b \le c$ then $\Delta(c,a) \le \Delta(c,b)$
\item if $a \le b$, then $\Delta(b,a) = b$ if and only if $a = b$
\item if $a < b$, then $\Delta(b,a) = b$ or $\Delta(b,a) \join b = \emptyset$
\end{enumerate}
\end{coro}

\begin{proof}
The result follows from \cite[Proposition 2.13]{MCL} and that $\De$ factors through the quotient map.
\end{proof}

We can see that $\Delta$ is order preserving (2,4,5), and it acts as the identity on the diagonal elements (1) of the product $M \times M$. In addition, it retains a notion of a local complement (6,7). 

Lastly, we look at the logical connective of implication.  
\begin{defin}
Let $\overline{M}$ be a multi-cubic lattice, $a, b\in \overline{M}$. Define implication by $b \r a = \Gamma(a) + X_{\sigma(a) \cup \sigma(b)^c}$ \cite[Definition 2.14]{MCL}.
\end{defin}

\begin{prop}\label{multicubes are implication algebras}
Multi-cubic lattices form implication algebras \cite[Lemma 2.15]{MCL}.
\end{prop}

Using the above result, we only need to show that implication is well defined on the quotient in order to form an implication algebra. 

\begin{lemma}\label{implication is well defined}
Implication is well defined when considered on the critical multi-cubic lattice.
\end{lemma}

\begin{proof}
 As the definition of implication is in terms of the critical element of a, we have that $b \r a =  b \r \Gamma(a)$. Therefore it is sufficient to show $b \r a = \Gamma(b) \r a$. 
 We note that for all $c \in M$ such that $\Gamma(c) = \Gamma(b)$, we have that $X_{\sigma(c)} = X_{\sigma(b)}$, and equivalently  $X_{\sigma(c)}^c = X_{\sigma(b)}^c$. Therefore the equality follows. 
\end{proof}

\begin{theorem}
Critical multi-cubic lattices are implication algebras with implication defined by  $b \r a = a + X_{\sigma(a) \cup \sigma(b)^c}$.
\end{theorem}

\begin{proof}
This follows directly from Proposition \ref{multicubes are implication algebras} and Lemma \ref{implication is well defined}.
\end{proof}

In summation, 
\begin{remark}\label{logic summary}
A critical multi-cubic lattice $M(\meet, \join, \De, \r)$ is an implication algebra, with a weak form of complementation and propositions faithfully embedded as projections onto a Hilbert space.
\end{remark}

We believe that this is the beginnings of a propositional quantum logic for higher spin qudits. The relation to qudits will be made clear in the following section. 
\section{Operator Algebras of a Critical Multi-Cubic Lattice}
Now that have an embedding of the critical multi-cubic lattice into a Hilbert space, we can consider the representation of its automorphism group. From this, we obtain the generalized Pauli matrices and the quantum Fourier transform.  To our knowledge, this is the first time, that these operators have acted on an infinite system of qudits.

\begin{defin}
Let M be a critical multi-cubic lattice such that  where $M \subseteq HL$ as in Theorem \ref{embedding multi-cubic}, $C$ be coatoms of $M$, and $A$ be the atoms of $M$. We denote the projection operator onto $c \in C$ as $p_{c}$, and the projection operator onto $a \in C$ as $p_{a}$.
\end{defin}

\begin{prop}\label{unitary representation multi-cubic}
Let $M$ be an $|I|$ critical multi-cubic lattice over $\Z_{2k+1}$. Then there exists a unitary representation $\rho: Aut(M) \r B(H)$ where $H$ is constructed in Theorem $\ref{embedding multi-cubic}$. 
\end{prop}

\begin{proof}
For any $\phi \in Aut(M)$, we equivalently consider $\phi \in Per_{Aut(\Z_{2k+1})} C$ by Lemma \ref{automorphism isomorphic to permutation}. We use coatomicity of the critical multi-cubic lattice to define an action on the atoms of $M$. As the atoms of M form an orthonormal basis of $H$, we can then consider the linear extension to obtain $\rho(\phi)$. Since $\rho(\phi)$ sends an orthonormal basis to an orthonormal basis, we have that it must be unitary.  
\end{proof}

Ultimately, we have made a choice in our construction of $\rho$. This is because we have an issue of ambiguity when translating actions on projection operators to a given subspace. As an example, $\pm I$ both define exactly the same action the atoms of $M$ when considered as unitary similarity transformations of projection operators. We now discuss an equivalent way of defining $\rho$ that highlights this. Recall that operators $p_a$ are rank one projections onto some vector forming an orthonormal basis of $H$. From another perspective, every atom is an infinite tensor product of exactly one element, $a_i \in H_i$ of our original choice of orthonormal basis of $H_i$ for each $i \in I$. In order to reduce ambiguity, for each $p_a$, and a given permutation $\phi: \{p_a\} \r \{p_a\}$, we chose the vector $a = \otimes_{i \in I} a_i$ as an equivalence class representative of vectors $\{h \in H: p_a = h \ran \lan h\} = \{z a: z \in \partial B_1(0) \subseteq \Co\}$, so $\rho(\phi): H \r H$ is defined by the map $a \mapsto \phi(a)$.

As a consequence of this choice, we have the following. 

\begin{prop}\label{multicube choice of representation}
Let $M \subseteq HL$ as in Theorem \ref{embedding multi-cubic} and $U\in W^\ast(\rho(Aut(\Z_{2k+1}))'$ be unitary. There exists a unitary $V \in \rho(Aut(M))$ such that $Ad_U = Ad_V : M \r M$ and $U = VS$ for $S \in W^\ast(\{p_c\}_{c \in C}) \cap W^\ast(\rho(Aut(\Z_{2k+1})))'$. 
\end{prop}

\begin{proof}
If $U \in W^\ast(\rho(Aut(\Z_{2k+1}))'$, then $Ad_U \in \rho(Aut(M))$ by Lemma \ref{automorphism isomorphic to permutation}. Now let $V = \rho(Ad_U) \subseteq W^\ast(U_\De)'$. Then $Ad_{V^\ast} = Ad_V^{-1}$, so $Ad_{UV^\ast}|_{M} = Ad_I|_{M}$. As the action of inner automorphism stabilizes $M$, $UV^\ast \in W^\ast(\{p_c\}_{c \in C})'$, so there exists $S \in W^\ast(\{p_c\}_{c \in C})'  = W^\ast(\{p_c\}_{c \in C})$ \cite[Proposition 3.1.15]{cubic} such that $U = VS$. Furthermore, $S = UV^\ast$, and we conclude that $S \in W^\ast(\rho(Aut(\Z_{2k+1})))'$ as well. 
\end{proof}

\begin{prop}\label{commutator of base ring}
$W^\ast(Aut(\Z_{2k+1}))' = W^\ast(\rho(Per_{Aut(\Z_{2k+1})} C), W^\ast(\{p_c\}_{c \in C}) \cap W^\ast(Aut(\Z_{2k+1}))')$.
\end{prop}

\begin{proof}
This proof is a direct application Proposition \ref{multicube choice of representation}.
\end{proof}

As an interesting side note, when reduced to the finite case, the following corollary can be viewed as generalization of the group theoretic result that $Z(B_n)$ is equal to the identity and the antipodal map. 

\begin{coro}
Let $M$ be a critical multi-cubic lattice over $\Z_{2k+1}$. Then $Z(\rho(Aut(M))) = \rho(Aut(\Z_{2k+1}))$.
\end{coro}

\begin{proof}
We only show one containment in  Proposition \ref{wreath product representation}. The reverse containment follows by Proposition \ref{commutator of base ring} and that $\rho(Aut(M)) \cap W^\ast(\{p_c\}_{c \in C}) = I$.
\end{proof}

We now reconstruct the relevant matrix unit structure of $B(H)$ in terms of critical multi-cubic lattice automorphisms. 

\begin{lemma}\label{isomorphism of B(H) multi-cubic}
Let $M$ be an $|I|$ critical multi-cubic lattice over $\Z_{2k+1}$, $C_\alpha$ be the coatoms of $M$ for a fixed index $\alpha \in I$, and $H$ be constructed in the manner of Theorem \ref{embedding multi-cubic}, then $B(H) \cong M_{2k}(B)$, where $B$ is the mutual commutant of a set of matrix units.
\end{lemma}

\begin{proof}
Let $(\cdot)$ denote the respective element in the permutation group contained in $M_{2k}(\Z_2)$ represented in the standard basis, and $C_\alpha$ denote the coatoms for a fixed index $\alpha \in I$. Now we claim the following matrix units form matrix units of $B(H)$. For $i \in C_\alpha$: 
\begin{align*}
    e_{ii} &= p_c \\
    e_{ij} &= e_{ii} (ij) \\
    e_{ji} &= (ij) e_{ii}
\end{align*}

We can directly compute that $\sum_{i \in C_\alpha} e_{ii} = I$, $e_{ij} = e_{ji}^\ast$ as permutation group is subgroup of the unitary group, and $e_{ij}e_{kl} = \delta_{jk}e_{il}$. Therefore, $B(H) \cong M_{2k}(B)$, where $B$ is the commutant of the matrix units, see Lemma 4.27 of \cite{alfsen3}.
\end{proof}

\begin{defin}\hypertarget{representation definition}{}
Let H be constructed as in Theorem \ref{embedding multi-cubic}. We define the unitary representation $\rho_i: C_{S_{2k}}(Aut(\Z_{2k+1})) \r B(H)$ as elements of the permutation group in $M_{2k}(\Z_2)$ of the $i$th index of the tensor product defining $H$. 
\end{defin}

\begin{lemma}\label{automorphism containment}
With the conditions of Lemma \ref{isomorphism of B(H) multi-cubic}, $W^\ast(\rho_i(C_{S_{2k}}(Aut(\Z_{2k+1}))), \{p_{C_i}\}_{l=1}^{2k}) \subseteq W^\ast(\{e_{ij}\}_{i,j =1}^{2k})$.
\end{lemma}

\begin{proof}
We claim $W^\ast(\rho_i(C_{S_{2k}}Aut(\Z_{2k+1})), \{p_{C_i}\}_{l=1}^{2k}) \subseteq W^\ast(\{e_{ij}\}_{i,j =1}^{2k})$.

The projections onto the appropriate coatoms are the diagonal elements by construction. In place of $\De$ enforcing the conditions to be a cubic lattice, we have that the module conditions of the critical multi-cubic lattice must be preserved. 

We identify $\phi \in C_{S_{2k}}(Aut(\Z_{2k+1})))$ with $\sigma_{\phi} \in S_{2k} \subseteq M_{2k}(\Z_2)$ represented by the permutation group. Thus, any element of $\rho_i(C_{S_{2k}}(Aut(\Z_{2k+1})))$ is a linear combination of matrix units. 
\end{proof}

Note that our previous results of \cite{cubic} only fully generalize to a particular subset of critical multi-cubic lattices. 

\begin{lemma}\label{B is mutual commutant}
With the conditions of Lemma \ref{isomorphism of B(H) multi-cubic}, $B = W^\ast(\rho_i(C_{S_{2k}}(Aut(\Z_{2k+1}))), \{p_{C_i}\}_{l=1}^{2k})'$ if and only if $2k+1$ is prime. 
\end{lemma}

\begin{proof}
It is sufficient to consider when $W^\ast(\{e_{ij}\}_{i,j =1}^{2k}) \subseteq W^\ast(\rho_i(Aut(\Z_{2k+1})), \{p_{C_i}\}_{l=1}^{2k})$, which is equivalent to the case when the action of $\rho_i(C_{S_{2k}}(Aut(\Z_{2k+1})))$ on $\{e_{ii}\}_{i = 1}^{2k}$ generates $\{e_{ij}\}_{i,j = 1}^{2k}$. This occurs exactly when $C_{S_{2k}}(Aut(\Z_{2k+1}))$ acts transitively on $\Z_{2k+1}- \{0\}$ with our relabeling of Theorem \ref{isomorphism of permutations and centralizer}. 

Let $e \ne \sigma \in Aut(\Z_{2k+1})$. Consider $C_{S_{2k}}(\sigma) \cong \Pi_{j=1}^l (\Z_{j} \wr S_{N_j})$. The action of $\Pi_{j=1}^l (\Z_{j} \wr S_{N_j})$ on $\sigma$ must preserve the cycle type of $\sigma$. Let $1 \le i < j \le 2k$, then the action of $\Pi_{j=1}^l (\Z_{j} \wr S_{N_j})$ can map $i$ to $j$ only if $i$ and $j$ are both in cycles of the same length. Thus, if $\Pi_{j=1}^l\Z_{j} \wr S_{N_j}$ acts transitively, $\sigma$ must have a cycle decomposition of $m$ cycles of length $\frac{2k}{m}$ where $m$ divides $2k$. On the other hand, these cycles are the orbits of the action $\langle\sigma \rangle$ on $\Z_{2k+1}$ disregarding the orbit of $e \in \Z_{2k+1}$. For $a \in \Z_{2k+1}$, if $|a| = 2k+1$, then $|orb(a)| = \phi(a)$, where $\phi$ denotes the Euler totient function. If $|a| < 2k+1$, then $|orb(a)| = \phi(2k+1)/\phi(d)$ for some $1 < d \,|\, (2k+1)$. As $2k+1$ is odd, $d > 2$ and $\phi(d) > 1$. Therefore, $\phi(2k+1)/\phi(d)) \ne \phi(2k+1)$, so $2k+1$ must be prime. 

Conversely, if $2k+1$ is prime, recall that $C_{S_{2k}}(Aut(\Z_{2k+1})) = C_{S_{2k}}(\sigma) \cong \Z_{2k}$ whose action as a $2k$ cycle in $S_{2k}$ is transitive on $\{a: 1 \le a \le 2k\}$. 
\end{proof}

We generalize the Hadamard matrix for a critical multi-cubic lattice. Let $M$ be an $|1|$ critical multi-cubic lattice over $\Z_{2k+1}$ where $2k+1$ is prime. Let $\Z_{2k} \cong \rho(C_{2k}(Aut(\Z_{2k+1})))$ be generated by the unitary $X_{2k}$. As $X_{2k}$ is unitary, there exists a unitary $U_{2k}$ such that $U_{2k}^{\ast} X_{2k}U_{2k} = D_{2k}$, where $D_{2k}$ is the diagonal matrix of the roots of $2k$ roots of unity in counterclockwise order. In the case where $2k+1 = 3$, $X_{2k} = X$, and $U_{2k} = H$, where $X$ is the Pauli gate, and $H$ is the normalized Hadamard gate.

\begin{defin}\label{quantum fourier transform definition}\hypertarget{quantum fourier transform definition}{}
Let $M$ be an $|I|$ critical multi-cubic lattice over $\Z_{2k+1}$ where $2k+1$ is prime. Define $U_H = \otimes_{i \in I} U_{2k}$,
\end{defin}

\begin{exmp}\label{representation of generalized pauli matrices}
As noted if $M$ is an $|1|$ critical multi-cubic lattice over $\Z_{3}$, then $U_{2k}$ is the Hadamard matrix up to a normalization constant. If $M$ is an $|1|$ critical multi-cubic lattice over $\Z_{5}$, then
\[U_{2k} = \frac{1}{2}
\begin{bmatrix}
 1 & 1 & 1 & 1\\
 1 & i & -1 & -i \\
 1 & -1 & 1 & -1 \\
 1 & -i & -1 & i \\
\end{bmatrix}, 
X_{2k} = \begin{bmatrix}
    0 & 1 & 0 & 0 \\
    0 & 0 & 1 & 0 \\
    0 & 0 & 0 & 1 \\
    1 & 0 & 0 & 0\\
\end{bmatrix}\text{, and } 
D_{2k} = \begin{bmatrix}
    1 & 0 & 0 & 0 \\
    0 & i & 0 & 0 \\
    0 & 0 & -1 & 0 \\
    0 & 0 & 0 & -i \\
\end{bmatrix}
.\]
In general up to a normalization constant, $\left[U_{2k}\right]_{ij} = \omega_j^{i-1}$ where $\omega_j$ is $j$th element of the $2k$ roots of unity with counterclockwise ordering starting at 1.  We recognize $U_{2k}$ as the quantum Fourier transform, $X_{2k}$ as the shift matrix, and $D_{2k}$ as the clock matrix.
\end{exmp}

Originally introduced by \cite{nonions}, it is well known that these matrices form the framework of quantum mechanical dynamics in finite dimensions. Furthermore, $X_{2k}$ and $D_{2k}$ are known as Generalized Pauli matrices, which are a representation of the respective Heisenberg-Weyl group \cite{A_Vourdas_2004}. 

\begin{defin}
Let $X_{(2k)_i} = \otimes_{j \in I} A_j$, where $A_i = X_{2k}$ and $A_j = I_{2k}$ for all $j \ne i$. Similarly, let $D_{(2k)_i} = \otimes_{j \in I} A_j$, where $A_i = D_{2k}$ and $A_j = I_{2k}$ for all $j \ne i$.
\end{defin}
    
\begin{prop}\label{unitrary similarity of coatoms contains X}
Let $M$ be an $|I|$ critical multi-cubic lattice over $\Z_{2k+1}$ where $2k+1$ is prime and $C$ be the coatoms of $M$. Then $W^\ast(\rho(C_{S_{2k}}(Aut(\Z_{2k+1}))) \subseteq W^\ast(\{X_{(2k)_i}\}_{i\in I}) \subseteq W^\ast(\{U_Hp_c U^\ast_H\}_{c \in C}$.
\end{prop}

\begin{proof}
We first show $W^\ast(\rho(C_{S_{2k}}(Aut(\Z_{2k+1}))) \subseteq W^\ast(\{X_{(2k)_i}\}_{i\in I})$. Since $2k+1$ is prime, $C_{S_{2k}}(Aut(\Z_{2k+1})) = Aut(\Z_{2k+1}) \cong \Z_{2k}$, so we reduce to the case of the representation of the cyclic group $\rho(\Z_{2k})$, which is a subset of the standard representation of the permutation group for our chose basis of Theorem \ref{embedding multi-cubic}. As $\rho(\Z_{2k}) = \otimes_{i\in I} X_{(2k)_i}$ by construction, we have shown the first containment. 

Now we show $W^\ast(\{X_{(2k)_i}\}_{i\in I}) \subseteq W^\ast(\{U_Hp_c U^\ast_H\}_{c \in C}$. We have $D_{(2k)_i} = \sum_{j=1}^{2k} \omega_j p_{j}$ where $\{\omega_j\}$ are the $2k$ cyclic roots of unity and $p_j$ are the rank one projections onto the basis used to construct $H_i$ in Theorem \ref{embedding multi-cubic}, so that $D_{(2k)_i}$ is diagonal in our choice of basis. As all $p_j$ are atoms of $M$, $D_{(2k)_i} \subseteq W^\ast(\{p_c\}_{c \in C})$ by coatomisticity. Lastly, we conclude that $X_{(2k)_i} = U_H D_{(2k)_i} U_H^\ast \subseteq W^\ast(\{U_Hp_c U^\ast_H\}_{c \in C})$.
\end{proof}

\begin{prop}\label{X contains centralizer}
Let $M$ be an $|I|$ critical multi-cubic lattice over $\Z_{2k+1}$ where $2k+1$ is prime and $C$ be the coatoms of $M$. Then 
$W^\ast(\{U_Hp_c U^\ast_H\}_{c \in C} \subseteq W^\ast(\{\rho_i(C_{S_{2k}}(Aut(\Z_{2k+1}))\}_{i\in I})$.
\end{prop}

\begin{proof}
It is sufficient to reduce to an arbitrary $i \in I$. By Proposition \ref{atomistic multi-cubic lattice}, we need only show that $W^\ast(X_{2k_i}) =  W^\ast(\{U_H p_{ij} U_H\}_{j = 1}^{2k})$, where $p_{ij}$ is the projection onto $jth$ atom forming an orthormal basis of $H_i$ used to construct $H$ in Theorem \ref{embedding multi-cubic}.

We have already shown one containment, so it sufficient to show that the vector spaces have equal rank. This follows as  $rank(W^\ast(X_{2k_i}) = 2k$, as we have a matrix with distinct eigenvalues, so $X_{2k_i}$ has minimal polynomial equal to its characteristic polynomial, $x^{2k} - 1$.  W$^\ast(\{U_H p_{ij} U_H\}_{j = 1}^{2k})) = 2k$ as well since it is just a unitary transformation of $2k$ orthogonal projections. 
\end{proof}

\begin{lemma}\label{reprentation of cyclic group equal hadamard transformation}
Let $M$ be an $|I|$ critical multi-cubic lattice over $\Z_{2k+1}$ where $2k+1$ is prime and $C$ be the coatoms of $M$. Then 
$W^\ast(\{\rho_i(C_{S_{2k}}(Aut(\Z_{2k+1})))\}_{i \in I}) = W^\ast(\{U_Hp_c U^\ast_H\}_{c \in C}$ if and only $2k+1$ is prime.
\end{lemma}

\begin{proof}
If $2k+1$ the result follows by Proposition \ref{unitrary similarity of coatoms contains X} and Proposition \ref{X contains centralizer}. 

Assume that $2k+1$ is not prime and $W^\ast(\rho_i(C_{S_{2k}}(Aut(\Z_{2k+1}))_{i \in I}) = W^\ast(\{U_Hp_c U^\ast_H\}_{c \in C}$. Specifically this implies that $C_{S_{2k}}(Aut(\Z_{2k+1}))$ contains a $2k$ cycle. In addition, $C_{S_{2k}}(Aut(\Z_{2k+1}))$ is an abelian centralizer and therefore a maximal abelian subgroup of $S_{2k}$. Similarly, $2k$ cycles are abelian subgroups equal to their centralizer in $S_{2k}$, so they are maximal abelian subgroups. Then $C_{S_{2k}}(Aut(\Z_{2k+1}))$ contains a $2k$ cycle if and only if $C_{S_{2k}}(Aut(\Z_{2k+1})$ is equal to a $2k$ cycle. This contradicts that $Aut(\Z_{2k+1}) \le C_{S_{2k}}(Aut(\Z_{2k+1})$ as either $Aut(\Z_{2k+1}) = \{1\}$ or $Aut(\Z_{2k+1}) = C_{S_{2k}}(Aut(\Z_{2k+1})$. The former is false as $|Aut(\Z_{2k+1})| = \phi(2k+1) > 1$ for all $k \ge 1$, and the latter holds if and only if $2k+1$ is prime.




\end{proof}

Our final theorem can be viewed as a generalization of the result that generalized Pauli matrices form an orthonormal basis. 

\begin{theorem}\label{B(H) multi-cubic isomorphism}
Let $M$ be an $|I|$ critical multi-cubic lattice over $\Z_{2k+1}$ where $2k+1$ is prime, $C$ be the coatoms of $M$, and $H$ be constructed as in Theorem \ref{embedding multi-cubic}. \\ Then $B(H) = W^\ast(\{U_Hp_cU^\ast_H\}_{c \in C}, \{p_c\}_{c \in C}) = W^\ast(\{\rho_i(C_{S_{2k}}(Aut(\Z_{2k+1})))\}_{i \in I}, \{p_c\}_{c \in C})$ if and only if $2k+ 1$ is prime.
\end{theorem}

\begin{proof}
By Lemma \ref{reprentation of cyclic group equal hadamard transformation}, the second equality holds if and only if $2k+1$ is prime. 
Now assume that $2k+1$ is prime, and consider a coatom in $p_i \in \{Up_cU^\ast\}_{c \in C}$ and $q_i \in \{p_c\}_{c\in C}$ for some fixed index $i \in I$. Then $p_i \meet q_i = \lim_{n \r \infty} (p_iq_ip_i)^n  = \lim_{n \r \infty} (\frac{1}{2k}p_i)^n = 0$. By construction, any atom $a \in W^\ast(\{Up_cU^\ast\}_{c \in C})$, $a$ is bounded by a coatom for all $i \in I$, so we assume without loss of generality that $a \le p_i$, and by symmetry we assume $b \le q_i$. Then $a \meet b \le p \meet q = 0$. Therefore the atomistic Boolean lattice of projections associated with $\{U_Hp_cU_H^\ast\}_{c \in C}$ and $\{p_c\}_{c \in C}$ have distinct sets of atoms.
By atomisticity, $W^\ast(\{p_c\}_{c\in C})$ and $W^\ast(\{U_Hp_cU_H^\ast\}_{c\in C})$ are abelian von Nuemann algebras whose only common projections are $0$ and $I$, so their intersection is $\Co I$ by \cite{stonean}. 
\end{proof}
Lastly, we have a result about the possible transitions of the critical multi-cubic lattice using only gates associated with the automorphism group.  

\begin{coro}
Let $M$ be an $|I|$ critical multi-cubic lattice over $\Z_{2k+1}$. The action of $Aut(M)$ acts transitively on the atoms if and only if $2k+1$ is prime. 
\end{coro}

\begin{proof}
As $Aut(M) \cong C_{S_{2k}}(Aut(\Z_{2k + 1})) \wr S_I$, we need only show that $C_{S_{2k}}(Aut(\Z_{2k + 1}))$ acts transitively on the coatoms on fixed index $i \in I$. As the action is just standard modular multiplication, we apply $C_{S_{2k}}(Aut(\Z_{2k + 1}))$ acts transitively on $\Z_{2k+1} - \{0\}$ if and only if $2k+1$ is prime. By coatomisticity of the critical multi-cubic lattice, we conclude the result. 
\end{proof}



Our original interest in the critical multi-cubic lattice began as a generalization of the cubic lattice. Fortuitously, as in \cite{cubic}, the appropriate representation of the critical multi-cubic lattice created a framework for infinite systems of quantum gates, in this case qudits. In order to do so, we have lost the Boolean-adjacent properties of the cubic lattice namely axiom 2, and therefore a nice characterization of the logic that followed. An interesting future pursuit would be to more fully discuss the implicative structure of the critical multi-cubic lattice. Another pursuit is to discover an intermediate structure between the generality of the critical multi-cubic lattice and the Boolean-like logic of the cubic lattice.

In order to highlight the utility and importance of moving between a logical formulation and algebraic formulation, we conclude with a quote by J. Sylvester: \textquote{The application of Algebra to Logic is now an old tale$-$the application of Logic to Algebra marks a far more advanced stadium of the human intellect,} \cite{nonions}.

\printbibliography[
heading=bibintoc,
title={References}
]
\end{document}